# Comment on decoherence by time dilation

H. Dieter Zeh (www.zeh-hd.de) - arxiv:1510.02239v2

In a recent publication, M. Pikovski et al. proposed a novel, purely internal decoherence mechanism that is based on differences in proper times along different paths of the center of mass of a mesoscopic system.[1] Although this mechanism is negligible compared to normal (collisional) decoherence by the environment for all realistic systems, and merely represents "virtual decoherence" (in being reversible similar to a spin echo if other sources of decoherence are neglected), it is an interesting effect from a fundamental point of view. So it has already aroused a number of (partly critical) comments by other authors.[2]

In contrast to attempts of postulating a collapse of the wave function by means of some fundamentally new "open systems dynamics" in order to describe classical behavior in quantum mechanical terms, decoherence is generally based on the assumption of global unitarity.[3] For consistency, this has then also to be assumed for quantum gravity, where it may lead to superpositions of different gravitational fields – as already discussed in 1957 by Richard Feynman by means of his example of a gravitational Schrödinger cat.[4] However, these superpositions would soon disappear from the point of view of all local observers through decoherence by matter variables, which thereby serve as an environment.[5] This leaves only some slight (in practice negligible) effective indeterminism of otherwise classical gravity. In quantized general relativity, though, the universal quantum state cannot depend on external time, but has instead to describe any time-dependence by means of an entanglement between the spatial metric and matter variables. Claus Kiefer has applied a Born-Oppenheimer approximation to the Wheeler-DeWitt equation in order to derive a Tomonaga equation (a relativistic Schrödinger equation of QFT with its many-fingered time) that is valid separately along all WKB trajectories of geometry.[6] The latter are robust under decoherence, and therefore essentially represent autonomous Everett branches of the WDW wave function, while quantum gravity is then not further required.

Such a Tomonaga equation (based on a classical spacetime geometry) is the general starting point for relativistic quantum theory in the Schrödinger picture, while the new decoherence model makes further simplifying assumptions. Here one calculates the non-relativistic quantum dynamics of all internal variables of a mesoscopic object along different center-of-mass trajectories with their different proper times. Upon relocalization, they must then possess different internal states, and in this way cause the new kind of decoherence for the center of mass wave function. This is claimed to explain the latter's classical behavior in terms of narrow, non-interfering wave packets representing an *effective ensemble* of quasi-classical paths or Everett branches (although a similar but quantitatively stronger result is already obtained by conventional decoherence). There is some dispute under what conditions precisely this calculation is correct (such as only for inertial or also for other reference systems).[2] I will here not enter this dispute, since the method seems to be valid in principle at least in *some* situations, provided the assumed different paths are sufficiently defined and distinct in this quantum setting.

The authors of the original paper furthermore assumed the internal degrees of freedom to consist of a large number of non-interacting oscillators that would form a thermal bath of finite temperature $T \neq 0$. Adler and Bassi[2] replaced this initial condition by a general *pure* internal state (which might also be a product of oscillator states, for example). They were then able to demonstrate that the new decoherence requires a non-vanishing energy variance of the pure state (as it must exist for a non-trivial time dependence) rather than non-vanishing temperature. So the new kind of decoherence disappears for all internal energy eigenstates rather than only for the ground state ($T = 0$). Even if one would assume the *total* internal system to be represented by a thermal bath that is classically explained by a lack of knowledge about the precise energy (an ensemble of energy eigenstates), this can *not* be responsible for a physical effect such as decoherence, since in this case only one member of the ensemble would represent individual reality. (This argument was first used by Eugene Wigner to exclude simple phase averaging in an ensemble, sometimes called "dephasing", as a possible solution of the measurement problem.) If a canonical density matrix for the *total internal mesoscopic system* is instead understood as representing previously existing entanglement with an unbounded environment (that is, an improper mixture), this would simply explain the decoherence arising by the new mechanism as a dynamical extension of this entanglement with the environment to the center of mass wave function – that is, indeed as "real" and environmental decoherence.[7] This confusion of quantum entanglement (responsible for decoherence) with incomplete "information" (ensembles of states) seems to form a very popular but subtle prejudice in quantum measurement theory.[8] The reduced density matrices of subsystems are insufficient to describe entanglement, as they would not even be able to predict the violation of Bell's inequality. It appears plausible to assume that *all* physically relevant "mixed states", in particular thermal baths, represent (mostly uncontrollable) physical entanglement in this sense, caused as *part of nonlocal reality* by earlier interactions, rather than somebody's incomplete information.

**References**

[1] I. Pikovski, M. Zych, F. Costa, and C. Brukner, Universal decoherence due to gravitational time dilation, Nature Physics **11**, 668 (2015)

[2] Y. Bonder, E. Okun, and D. Sudarsky, Can gravity account for the emergence of classicality?, arXiv:1507.05320; S. L. Adler and A. Bassi, Gravitational decoherence for mesoscopic systems, arXiv:1506.04414; C. Gooding and W. G. Unruh, Bootstrapping time dilation decoherence, Found. Phys. **45**, 1166 (2015); L. Diosi, Center of mass decoherence due to time dilation: paradoxical frame dependence, arXiv:1507.05828; I. Pikovski, M. Zych, F. Costa, and C. Brukner, Time dilation in quantum systems and decoherence: questions and answers, arXiv:1508.03296

[3] For an overview of its role in the interpretation of measurements, see H. D. Zeh, The strange (hi)story of particles and waves, arXiv:1304.1003, Sect. 4

[4] H. D. Zeh, Feynman's interpretation of quantum theory, Eur. Phys. J. **H36**, 147 (2011) – arXiv:0804.3348

[5] E. Joos, Why do we observe a classical spacetime?, Phys. Lett. **A116**, 6 (1986)

[6] C. Kiefer, Quantum Gravity, 3rd edn. (Cambridge 2012)

[7] See also the footnote on page 15 of version 14 of Ref. 3

[8] H. D. Zeh, John Bell's varying interpretations of quantum mechanics, arXiv:1402.5498